\documentclass[]{aa}
\usepackage[]{psfig}
\usepackage[mathscr]{eucal}
\usepackage[amsfonts]{}
\begin{document}

\title{Period-doubling events in the light curve of R~Cygni: 
evidence for chaotic behaviour}

\author{L. L. Kiss \and K. Szatm\'ary}

\institute{Department of Experimental Physics and Astronomical Observatory,
University of Szeged,
Szeged, D\'om t\'er 9., H-6720 Hungary}

\titlerunning{Chaotic behaviour of R~Cygni}
\authorrunning{Kiss \& Szatm\'ary}
\offprints{l.kiss@physx.u-szeged.hu}
\date{}

\abstract{
A detailed analysis of the century long visual light curve of the long-period
Mira star R~Cygni is presented and discussed. The data were collected from the
publicly available databases of the AFOEV, the BAAVSS and the VSOLJ. 
The full light curve 
consists of 26655 individual points obtained between 1901 and 2001. The light 
curve and its periodicity were analysed with help of the $O-C$ diagram,
Fourier analysis and time-frequency analysis. The results 
demonstrate the limitations of these linear methods.
The next step was to investigate the possible presence of low-dimensional
chaos in the light curve. For this, a smoothed and noise-filtered 
signal was created from the averaged data and with help of time delay embedding,
we have tried to reconstruct the attractor of the system. The main
result is that R~Cygni shows such period-doubling events that can be
interpreted as caused by a repetitive bifurcation of the chaotic 
attractor between a period $2T$ orbit and chaos. The switch between these two
states occurs in a certain compact region of the phase space, where the 
light curve is characterized by $\sim$1500-days long transients. 
The Lyapunov spectrum was computed for various embedding parameters 
confirming the chaotic attractor, although the exponents suffer from 
quite high uncertainty because of the applied approximation.
Finally, the light curve 
is compared with a simple one zone model generated by a third-order
differential equation which exhibits well-expressed period-doubling
bifurcation. The strong resemblance is another argument for 
chaotic behaviour. Further studies should address the 
problem of global flow reconstruction, including the 
determination of the accurate Lyapunov exponents and dimension.
}
 
\maketitle

\section{Introduction}

Red giants on the Asymptotic Giant Branch (AGB) of the Hertzsprung-Russell
diagram are pulsationally unstable and show characteristic light variations.
The traditional classification (Kholopov et al. 1985--88)
is based on two properties of the visual light
curve, the periodicity and full amplitude. Monoperiodic stars
with amplitudes larger than 2\fm5 are the Mira-type variables, smaller
amplitude or complex light curves place a star among the semiregular variables.
Even the monoperiodic Mira stars do not have strictly repeating light variations,
as there are apparently irregular cycle-to-cycle changes in amplitude and/or 
period. The overwhelming majority of 
related studies addressed the problem of period change, from the early 
works of Eddington \& Plakidis (1929), Sterne \& Campbell (1936) to
more recent studies of Isles \& Saw (1987), Lloyd (1989), Percy \& Colivas
(1999) and long series of papers by Koen \& Lombard (from Koen \& Lombard 1993 to 
Koen \& Lombard 2001). Besides a few detections of significant gradual 
period change (e.g. G\'al \& Szatm\'ary 1995, Sterken et al. 1999), 
most of the studies concluded that random period fluctuations dominate
the Mira light curves. Interestingly, the amplitude variations remained
quite unstudied, likely because of the nature of the only existing
data -- low precision visual observations. Very few investigations 
can be found on this issue, the most important ones include an analysis 
of observations by Canizzo et al. (1990) and a theoretical study by Icke 
et al. (1992). Both papers inspected the possible presence of chaos 
in long-period variables. While Canizzo et al. (1990) rejected the 
hypothetic low-dimensional chaos in the three stars studied,
Icke et al. (1992) argued that there is a lot of evidence for chaotic 
behaviour in the red giant variations. 

Variable stars are primary targets of nonlinear
analyses attempting to detect low-dimensional chaos in astrophysical systems.
The most successful detections are associated with pulsating 
white dwarfs (e.g. Goupil et al. 1988, Vauclair et al. 1989), 
W~Vir model pulsations (Buchler \& Kov\'acs 1987, Serre et al. 1996b)
and two RV~Tauri type stars, R~Sct (Koll\'ath et al. 1990, Buchler et al. 1996)
and AC~Her (Koll\'ath et al. 1998). A common feature is the 
period-doubling bifurcation, which occurs in many simplified 
model oscillators, too (e.g. Saitou et al. 1989, Seya et al. 1990, Moskalik
\& Buchler 1990). Most recently, Buchler et al. (2001) presented 
an overview of observational examples of low-dimensional chaos including
the first preliminary results for four semiregular stars
(SX~Her, R~UMi, RS~Cyg and V~CVn). The emerging view of irregular
pulsations is such that large luminosity-to-mass ratio strongly
enhances the coupling between the heat flow and the acoustic oscillation.
Consequently, the relative growth rates for relevant pulsation modes 
are of order of unity which is the necessary condition for chaotic
behaviour. Since it is naturally satisfied in the low-mass, high-luminosity
AGB stars, they are likely candidates for chaotic pulsations.

In this study, we analyse the visual light curve of the Mira star R~Cygni (=
HD 185456 = IRAS 19354+5005, spectral type S6/6e with Tc-lines, Jorissen et
al. 1998).  It is a bright, well-observed variable star with an average
period of $\sim$430 days and average light extrema at $\sim$7\fm0 and
$\sim$14\fm0. The light variation was discovered by Pogson in 1852 and the
available magnitude estimates date back to the late 19th century. It has
been studied by numerous authors (the ADS lists 179 references), mostly
spectroscopically. Wallerstein et al. (1985) pointed out the correlation
between brightness at maximum and interval from the previous cycle (fainter
maxima occur later than normal). R~Cyg was also included in a sample of 355
long period variables studied by Mennessier et al. (1997) who presented a
classification of the light curves of LPVs for a discrimination between
carbon- and oxygen-rich stars. To our knowledge, there is no further study
dealing with the light variation of this star.

The main aim of this paper is to present a thorough analysis of the
hundred-years long light curve. The paper is organised as follows. The
observations and data preparation are described in Sect.\ 2. An
intention to interpret the light curve with standard linear methods is
discussed in Sect.\ 3. The nonlinear analysis is presented in Sect.\ 4,
while concluding remarks and further directions are listed in Sect.\ 5.

\section{Observations and data preparation}

\begin{figure}
\begin{center}
\leavevmode
\psfig{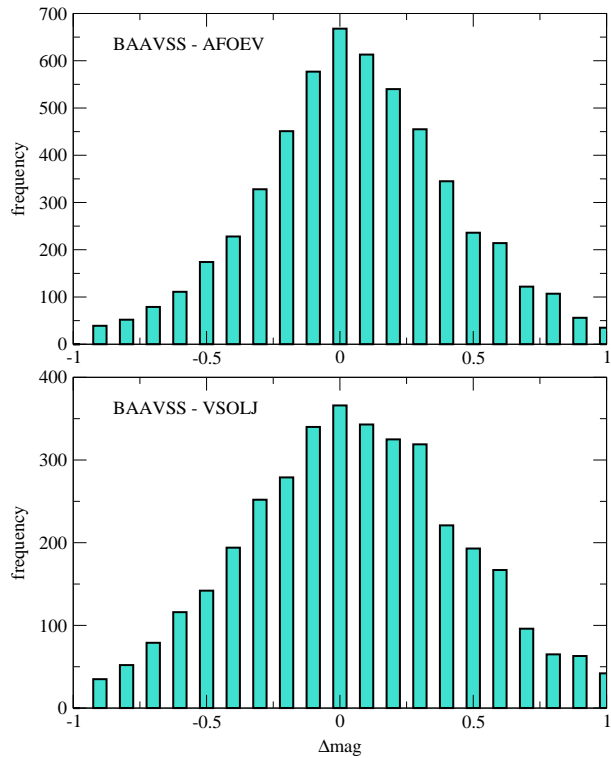}
\caption{The distribution of the magnitude differences computed for 
10-day means of the BAAVSS and AFOEV data (top) and the BAAVSS and AFOEV 
data (bottom). The maxima at 0 indicate good agreement between the 
different data.}
\end{center}
\label{fig1}
\end{figure}

\begin{figure}
\begin{center}
\leavevmode
\psfig{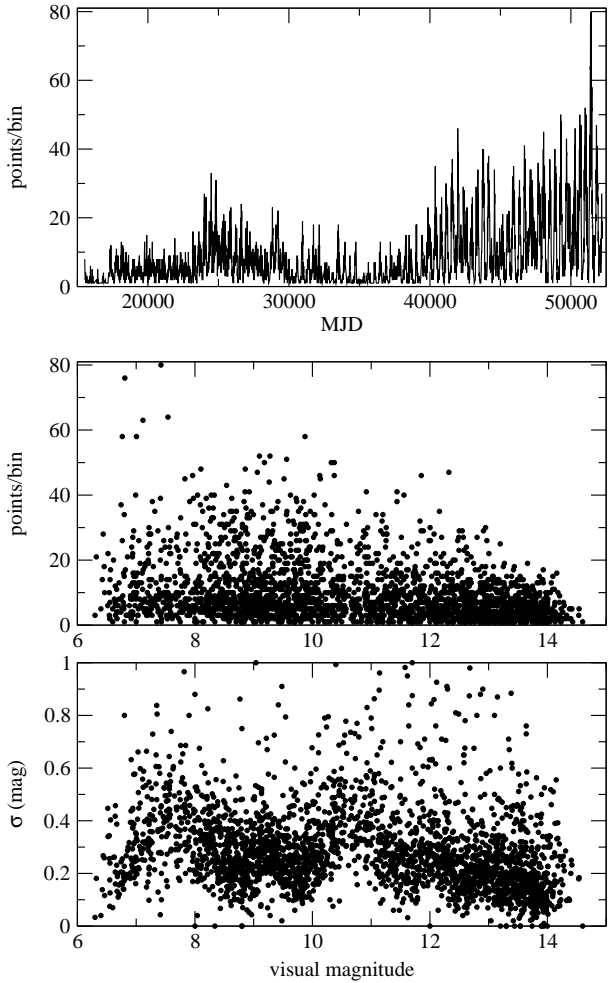}
\caption{{\it Top panel}: The evolution of number of points per bin
along the light curve. {\it Middle panel:} The same parameter against
the apparent magnitude. Note the slightly triangular distribution with
more data in the brighter states. {\it Bottom panel:} The mean deviation 
of estimates from the mean values (denoted with $\sigma$).}
\end{center}
\label{fig2}
\end{figure}

\begin{table}
\begin{center}
\caption{A summary of the analysed datasets (MJD=JD$-$2400000).}
\begin{tabular}{|lllr|}
\hline
Source & MJD(start) & MJD(end) & No. of points\\
\hline
AFOEV  & 22805 & 52181 & 8835\\
BAAVSS & 15513$^a$ & 51907 & 11024\\
VSOLJ  & 22597 & 51539 & 5888\\
VSNET  & 49718 & 52237 & 1105\\
\hline
Total: &  15513 & 52237 & 26852\\
Corrected: & 15513   & 52237 & 26655\\
\hline
\end{tabular}
\end{center}
$^a$ 8 points between MJD 11737 and 13411 were excluded.
\end{table}

Data were taken from the databases of the 
following organizations: the Association Fran\c caise 
des Observateurs d'Etoiles Variables (AFOEV\footnote{\tt 
ftp://cdsarc.u-strasbg.fr/pub/afoev}), the British Astronomical
Association, Variable Star Section (BAAVSS\footnote{\tt
http://www.britastro.org/vss/}) and the Variable Star Observers's
League in Japan (VSOLJ\footnote{\tt http://www.kusastro.kyoto-u.ac.jp/vsnet/gcvs}).
Since these data end in early 2001, the latest 
part of the light curve is covered with data obtained via the VSNET 
computer service.

The basic properties of the datasets are listed in Table\ 1. The 
three main sources are highly independent, as very few (less then 1\%) 
duplicated and no triplicated data points were found. The total number 
of estimates (duplicates counted twice) is 26852. Before merging the data 
we have checked the data homogeneity with a simple statistical 
test. We have calculated 10-day means for all of the three major data
sources (BAAVSS, AFOEV, VSOLJ) and compared their values in the 
overlapping regions. Since the British data are the longest one, we computed 
magnitude differences in sense of $\langle$BAAVSS$\rangle$ {\it minus} 
$\langle$AFOEV$\rangle$ and $\langle$BAAVSS$\rangle$ {\it minus}
$\langle$VSOLJ$\rangle$. Their distribution is plotted in Fig.\ 1. While the
AFOEV data result in a quite symmetric distribution, there is a
suggestion for the VSOLJ data for being slightly brighter in average than
the BAAVSS data. However, the highest peak at $\Delta$mag=0 and 
the symmetric nature for both distributions suggest
that there is no need to shift any of the data and thus, we simply merged
all data. After merging the original data files, we checked the deviant
points by a close visual inspection of the combined light curve. 
Almost 200 points were rejected and the finally adopted set 
contains 26655 visual magnitude estimates. 

The first step in the data processing was averaging the light curve using
ten-day bins. With the binning procedure some statistical findings were
revealed. First, we plotted the number of points per bin against time (Fig.\
2, top panel). Here one can find fairly high number of points per bin (with
steady increase in time) which is a necessary condition for calculating
accurate mean values. Second, the
distribution of the same parameter against the apparent magnitude (Fig.\ 2,
middle panel) shows that there are typically 5--10 points per bin even in
the faintest state, which is enough for good mean values. Finally, the mean
deviation of the individual points against the apparent magnitude (Fig.\ 2,
bottom panel) does not show significant trends, therefore, from a
statistical point of view, the calculated mean light curve can be regarded
quite homogeneous. The typical uncertainty of a single observation can be
thus estimated as $\pm$0\fm5. Consequently, the uncertainty of the calculated
mean values ($\approx\pm0\fm5/\sqrt{N_{\rm obs}}$) ranges from $\pm$0\fm08
to $\pm$0\fm17. The binned light curve is presented in Fig.\ 3. The fact,
that there are quite long subsegments, in which alternating maxima are
obviuously present, reminded us of the case of RV~Tauri type stars, i.e.
pulsating yellow giants with alternating minima. This alternation in 
R~Sct and AC~Her was interpreted as being caused by
low-dimensional chaos (Koll\'ath 1990, Koll\'ath et al. 1998) and that is
why we tackled the question of possible chaotic behaviour
in the light curve of R~Cyg. The light curve shape also closely resembles
that of some pulsating white dwarfs (e.g. Goupil et al. 1988), though 
with a characteristic time-scale of five order of
magnitudes larger.

Since we intended to study the light variation with tools of the nonlinear
time-series analysis, further preprocessing was done by smoothing and
interpolating the binned light curve. First, we fitted Akima splines (Akima,
1970) to the binned data to get an equidistant series of light curve points,
keeping the 10-day length (it means $\approx$43 points per cycle which is a
fairly good sampling). The spline fit was further smoothed with a Gaussian
weight-function, where we choose a wider (FWHM=20 days) Gaussian than in the
case of studies of semiregular variables (Kiss et al. 1999, Lebzelter \&
Kiss 2001). We have experimented with various FWHMs starting from 5 days up
to 30 days; the nonlinear analysis showed that much clearer images of the
attractor can be reached if the observational noise, as well as
remaining noise that is extraneous to the star (e.g. high-frequency
components due to low-amplitude, but high-dimensional jitter, see Buchler et
al. 1996) is smoothed out. The adopted 20 days provided well-smoothed signal
with no zig-zags in the light curve and avoided
strong amplitude decrease due to oversmoothing as in the case
of FWHM=30 days. The final result is a dataset \{$s(t_{\rm n})$\} that is
sampled at constant 10 days intervals. In Fig.\ 4 we show some typical
observed light curve segments with the smoothed and noise-filtered signal.
Obviously, some sharp features are missed in the latter data, however, as
will be presented later, the success of our nonlinear analysis supports our
data processing. In order to enable for the reader to study this interesting
star with other nonlinear methods, we make the original 10-day means and the
smoothed and noise-filtered data publicly available in electronic form at
CDS, Strasbourg\footnote{Corresponding data files can be found at CDS via
anonymous ftp to {\tt cdsarc.u-strasbg.fr (130.79.128.5)} or via {\tt
http://cdsweb.u-strasbg.fr/cgi-bin/qcat?J/A+/.../...}}

\begin{figure*}
\begin{center}
\leavevmode
\psfig{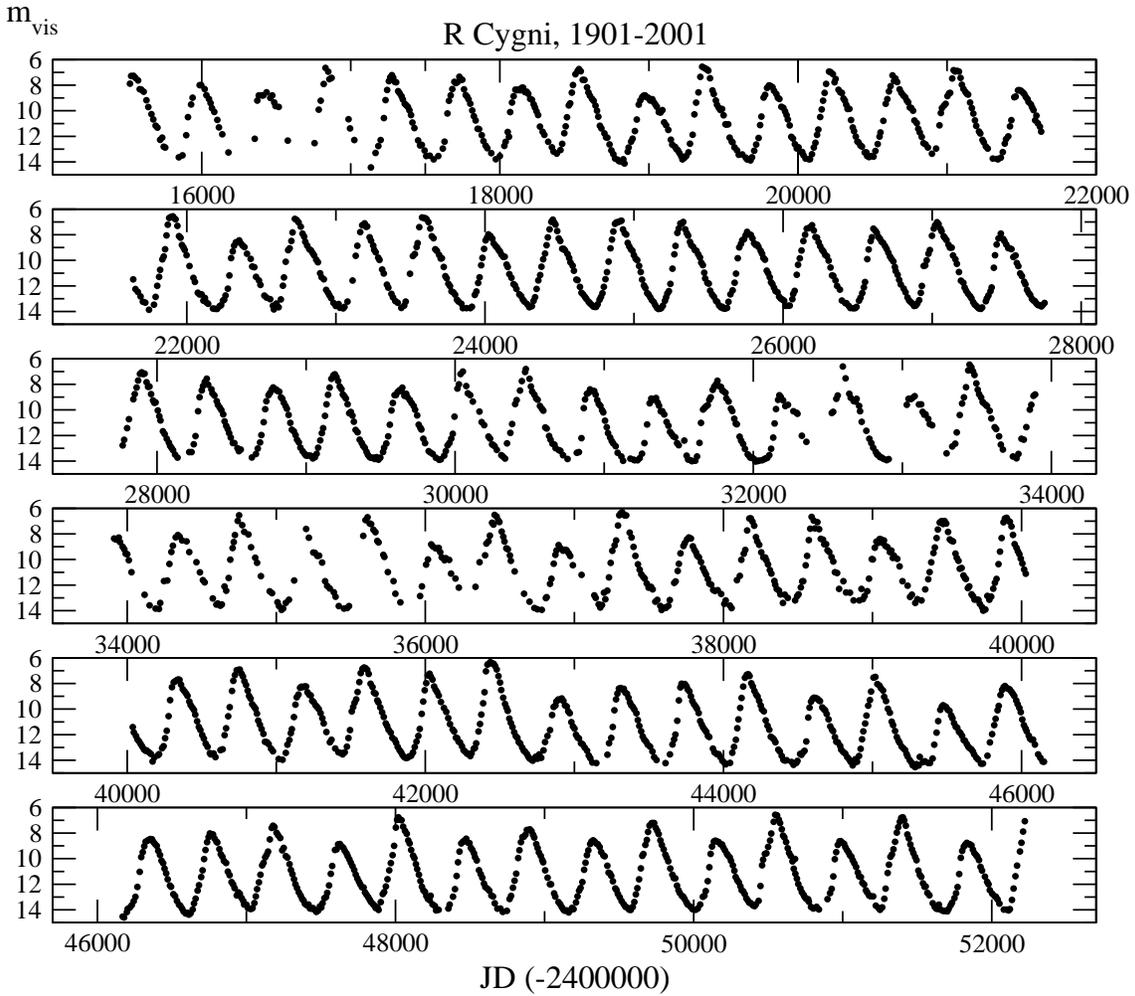}
\caption{The whole dataset of R Cygni (10-day means). Note the 
presence of alternating maxima, especially in the lowest two panels.}
\end{center}
\label{fig3}
\end{figure*}

\begin{figure}
\begin{center}
\leavevmode
\psfig{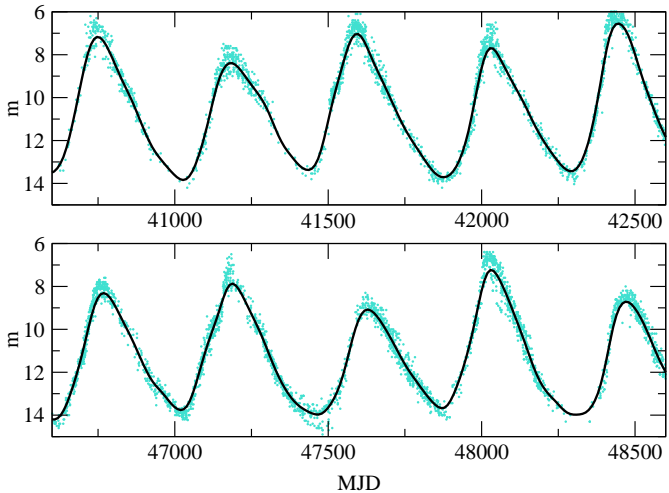}
\caption{Typical observed light curve segments (light dots)
with the smoothed and noise-filtered signal (solid line).}
\end{center}
\label{fig4}
\end{figure}

\section{Standard methods}

In this Section the period variation is examined with the $O-C$ diagram and
one of its descendants, then we show that Fourier-analysis fails to give a
full description of the light curve.  Furthermore, fairly strong subharmonic
components occur in the frequency spectrum which will be naturally included
in the adopted nonlinear description discussed in the next Section. Two
time-frequency methods are used to draw some constraints on the stability of
the frequency content.

\subsection{Period change}

Traditional variable star research studying single-periodic light variations
strongly relies on the $O-C$ method, although it has been shown in many 
cases that Mira stars usually do not fit the requirements of its
application. This is due to the large intrinsic period fluctuations
found in these variables (see, e.g. recent studies by 
Sterken et al. 1999, Percy \& Colivas 1999), but even white noise 
(Koen 1992) or low-dimensional chaotic systems (Buchler \& Koll\'ath 2001)
can generate curved $O-C$ diagrams leading to spurious detections of 
long-term period changes. Nevertheless, despite the many possible interpretations,
it might be interesting to show the results of this traditional method.

In order to construct the $O-C$ diagram we have determined all 
times of maxima from the binned light curve shown in Fig.\ 3. 
This was done by fitting low-order (3--5) polynomials to the selected
parts of the light curves around maxima. Typically, we selected $\pm$60--70
days around a given maximum (this means $\sim$13--15 points of the light
curve) and fitted polynomials with different orders chosen to give optimal
fit (as judged by a visual inspection of the data). The fitted interval
slightly changed with time as several cycles were covered only partially. 
Thanks to the extraordinarily good global light curve coverage, 
no cycle was lost between 1901 and 2001, thus
we could determine 86 epochs of maximum. A mean period of 428 days 
given by the Fourier analysis of the whole dataset (next subsection)
and the eighth epoch (MJD 18531) was used to construct the $O-C$ 
diagram plotted in Fig.\ 5. 

\begin{figure}
\begin{center}
\leavevmode
\psfig{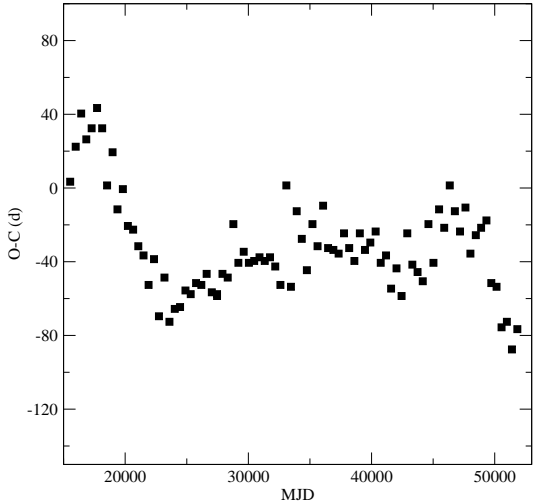}
\caption{The $O-C$ diagram of R~Cyg ($E_0$=MJD 18531, $P$=428 d).
The typical uncertainty of a given point is about $\pm$10 days.}
\end{center}
\label{fig5}
\end{figure}

Three different ``linear'' segments appear in the $O-C$ diagram
which could be explained by two period jumps at MJD 23500 and 
MJD 47200. However, little can be said about the astrophysical meaning
of these jumps. Following the work by Percy \& Colivas (1999), 
we performed the Eddington-Plakidis
test (Eddington \& Plakidis 1929) to estimate the amount of random 
period fluctuation $\epsilon$. The resulting relative fluctuation
$\epsilon/P$
is 0.01 that places R~Cygni close to the lower boundary of Fig.\ 2 in
Percy \& Colivas (1999), which means a relative stability of the 
mean period compared to other Mira stars of similar periods.
Our most important conclusion here is that we can exclude 
the star to be undergoing noticeable long-term evolution (due to, e.g., 
a thermal pulse deep in the stellar interior, Wood \& Zarro 1981)
over the one century of observations. The apparent period is shorter
at the very end of the dataset, but its variation does not seem to be 
a result of a slow period evolution. We note, that Koen \& Lombard
(2001) used simultaneously the epochs of maxima and minima to test
for period changes in long-period variables which is a more effective
approach than that of the $O-C$ diagram -- at least in those cases,
where only times of maxima and minima are available. In our case, the
use of the full light curve is strongly favoured instead of examining
any particular part of it.

\subsection{Fourier analysis}

\begin{figure}
\begin{center}
\leavevmode
\psfig{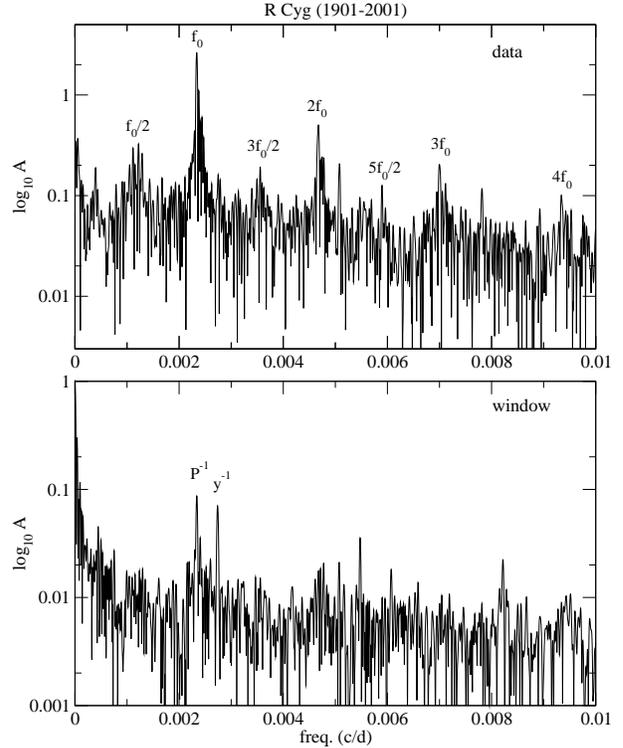}
\caption{The frequency spectrum of R~Cygni (top panel). Note the presence of 
subharmonic components up to 5/2 $f_0$. The window function shows 
two alias peaks, one is the 1-year alias due to seasonal gaps, 
while the other one at 0.00233 d$^{-1}$ (i.e. 1/430 d$^{-1}$) is caused by 
a few unobserved minima in the light curve.}
\end{center}
\label{fig6}
\end{figure}

\begin{figure}
\begin{center}
\leavevmode
\psfig{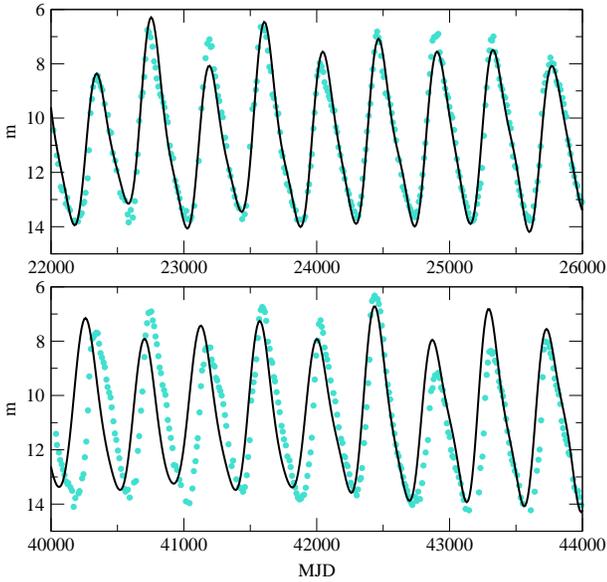}
\caption{A comparison of the 15 frequency fit with the interpolated
(top) and extrapolated (bottom) data (see text for details). 
The extrapolation obviously fails.}
\end{center}
\label{fig7}
\end{figure}

To investigate possible multiperiodicity, we have carried out a 
Fourier analysis with subsequent prewhitening steps. We have used Period98
of Sperl (1998) which also includes multifrequency least squares fitting of
the parameters. The studied frequency range was 0--0.01 d$^{-1}$ with a step
size of $1.8\cdot10^{-6}$. We have also determined the main period in all 6
subsets plotted in Fig.\ 3 to compare apparent period change with the
suggestions of the $O-C$ diagram.

The calculated amplitude spectrum of the whole light curve
is presented in Fig.\ 6, where 
the window function is shown, too. The primary peak is at 
$f_0$=0.002338 d$^{-1}$ ($P=428$ d). The structure of the
frequency spectrum is quite complex. One can find the integer harmonics
of the main frequency (2$f_0$, 3$f_0$, 4$f_0$), while subharmonic
components are present, too (3$f_0$/2, 5$f_0$/2). Note, that these 
subharmonic components are not introduced by the Fourier approach
through subsequent prewhitening with sine curves, as Fig.\ 6 shows 
the frequency spectrum before any prewhitening. Their presence is 
due to the alternating amplitude changes; this has been proven by a 
very simple test. We have created artificial light curves by repeating
{\it i)} a single cycle 50 times and {\it ii)} a double cycle with 
successive low and high amplitudes 25 times. Then the computed Fourier
spectra clearly marked the reason for the subharmonics.

The main problem is that instead of well-determined narrow peaks, 
there are closely separated groups
of peaks at the mentioned frequency values. The first 25 prewhitening
steps resulted in 8 components in the range 0.002249--0.002464 d$^{-1}$
($\approx f_0$), 4 components in the range 
0.004605--0.004781 d$^{-1}$ ($\approx 2f_0$),
6 components in the range 0.001007--0.001437 d$^{-1}$ ($\approx f_0/2$), 1 component
at 0.003558 ($\approx 3f_0/2$), 1 component at 0.007001 ($\approx 3f_0$) 
and 5 low-frequency components corresponding to the apparently irregular 
mean brightness variation. However, even the 25-frequency fit
leaves a residual scatter of 0\fm52 being much larger than the 
expected uncertainty of the points. Therefore, the fit, besides being
physically irrelevant, is still not perfect. 

Another test of the multiperiodic hypothesis was done as follows.
We have taken the first half of data, between MJD 15000 and 34000. This 
subset was Fourier analysed with the same procedure. The first 15 
frequencies yielded an acceptable fit of the whole subset. Then
this set of frequencies (and amplitudes and phases) was used to
predict the second half of data. As expected, the extrapolation
failed. This is demonstrated in Fig.\ 7, where subsets of the
halves are plotted with the interpolating and extrapolating 
fits. The extrapolated signal has obviously lost its predictive
power. Again, we have to reject the multiperiodic description
of the light curve.

The frequency grouping found in R~Cyg is exactly the 
same what as the one found for R~Sct by Koll\'ath (1990)
and AC~Her by Koll\'ath et al. (1998). In those cases detailed 
tests showed that multiple periodicity can be excluded as the reason 
for the high number of components in the Fourier spectrum. Furthermore,
such behaviour is characteristic for chaotic systems, e.g., the well-studied
R\"ossler oscillator (see Serre et al. 1996a and Buchler \& Koll\'ath 2001
for detailed discussions in astronomical terms).

The presence of subharmonic components is another interesting 
property. Similar subharmonics were found in a number of 
pulsating white dwarf stars in which they were interpreted 
as indications of period-doubling bifurcation (e.g. Vauclair et al. 1989). 
The $f_0$/2 subharmonic is
usually accepted indicator of period-doublings. This
subharmonic is fairly strong in the R\"ossler oscillator
(see, e.g., Fig.\ 3 in Serre et al. 1996a) or
other chaotic systems where pronounced period-doubling
occurs (a closely related example is the case of W~Vir hydrodynamic
models analysed by Serre et al. 1996b).

The separate analysis of 6 segments, each being $\sim$6000 days long
illustrates the essentially insignificant changes of the main period.  The
resulting periods are the following: MJD 15500--21600: 423$\pm$1.5 d; MJD
21600--27700: 429$\pm$2 d; MJD 27700--33900: 428$\pm$1 d; MJD 34000--40000:
428$\pm$0.5 d; MJD 40000--46100: 427$\pm$3.2 d; MJD 46100--52200: 422$\pm$1
d. This gives the same picture of the period change as the $O-C$ diagram
does: slightly shorter periods at the beginning and end of the data, between
them the mean period was constant. However, the given differences are
consistent with the random fluctuations of the period at 1\% level.

\subsection{Time-frequency analysis}

\begin{figure*}
\begin{center}
\leavevmode
\psfig{figure=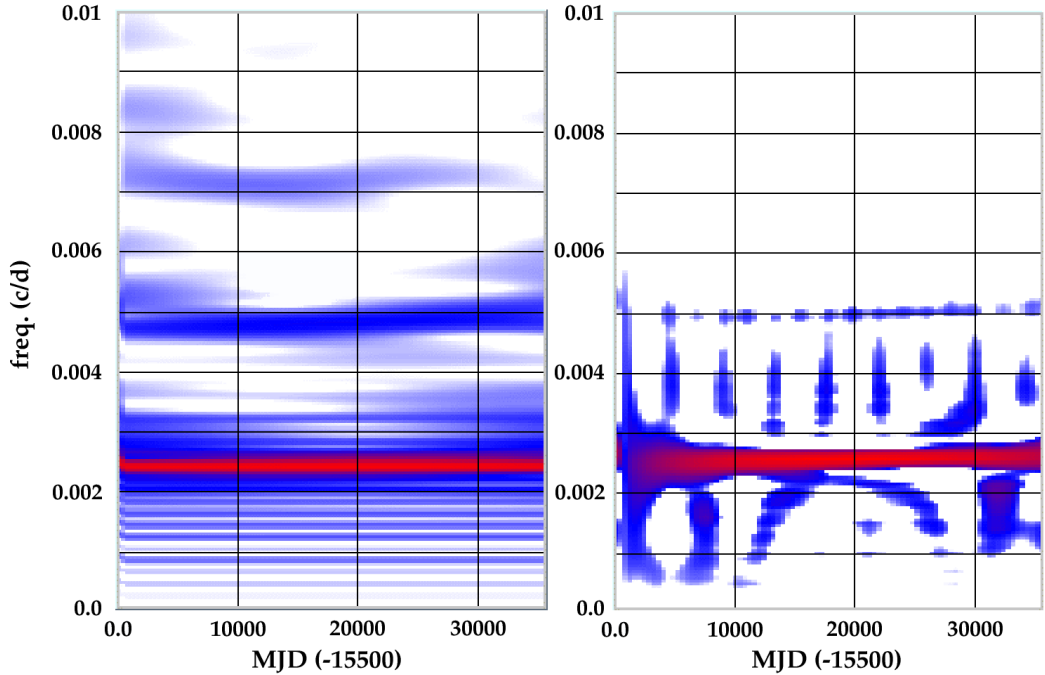,width=14cm}
\caption{The wavelet map (left) and Choi-Williams distribution (right) for R~Cyg.}
\end{center}
\label{fig8}
\end{figure*}

Modulations and sudden changes of the light curve (i.e. time-dependent
frequency content) can be studied with several time-frequency methods.
In variable star research, one of the widely used methods 
is the wavelet analysis (Szatm\'ary et al. 1994 and references 
therein). A more efficient method is the use of of Choi-Williams
distribution (CWD, Choi \& Williams 1989) which enhances the sharp features
in the time-frequency-amplitude space.

We present both the wavelet map and CWD in Fig.\ 8.
They were calculated with the software package TIFRAN (TIme FRequency
ANalysis) developed by Z. Koll\'ath and Z. Csubry at Konkoly Observatory,
Budapest. As expected, the main ridge at $f\approx0.0025$ d$^{-1}$ is 
the most obvious 
feature, while its harmonics are visible up to the third order.
Subharmonic components can be identified in the CWD and we also 
noticed a slight tilt of the ridges (most emphasized for $2f_0$) in
the last third of the data. The suggested period shortening around 
MJD 40000 (note the $-$15500 shift of the horizontal axes in Fig.\ 8)
is in good agreement with the results of the Fourier analysis.
Unfortunately, little can be inferred 
about the physical system behind the light variation. Time-frequency
analysis, similarly to Fourier analysis, is basically 
an interpolation method. There are strict limitations and 
real understanding needs a completely different approach. This
is provided by the methods of nonlinear time-series analysis.

\section{Nonlinear analysis}

During the last two decades, there has been a growing interest in   
nonlinear studies of variable stars. Several examples of seemingly 
irregular light curves were interpreted as results of low-dimensional
chaotic behaviour. Buchler and his co-workers published 
some fundamental papers dealing with either theoretical interpretations
or analyses of observations. There has been a number of excellent
papers and reviews in the astronomical literature on this 
topic (e.g. Perdang 1985, Serre et al. 1996a, Buchler et al. 1996,
Buchler \& Koll\'ath 2001), where the reader can find
definitions and descriptions of the analysing methods.

Hereafter we assume that the light curve is generated by a deterministic
dynamical sytem of a physical dimension $d$. In this section, we employ
various techniques to unfold the multidimensional structure of this system
in reconstructed phase spaces. The most important method is the time delay
embedding. We discuss some properties of the embedding and the reconstructed
attractor of the system, based on the smoothed and noise-filtered light
curve signal
$\{s(t_{\rm n})\}$. For this purpose, we have extensively used the TISEAN
package of Hegger et al. (1999), which is a publicly available software
package consisting of practical implementations of nonlinear time-series
methods\footnote{See also {\tt
http://www.mpipks-dresden.mpg.de/\~{}tisean}}. In the following we will
refer to some routines of this package (e.g. {\tt mutual}, {\tt
false\_nearest}, {\tt svd} etc.). Their use and basics are very well
documented at the cited website.

\subsection{Embedding parameters}
  
The first issue is the optimal choice of embedding parameters. As has been
discussed in, e.g., Buchler et al. (1996) different time delays
($\Delta$) and/or
embedding dimension ($d_{\rm e}$) may result in completely different phase
space reconstructions. In the case of R~Sct and AC~Her, a small range of
fraction of the formal period (5--20\%) yielded ``optimal'' results. We
applied the routine {\tt mutual} which calculates the time delayed mutual
information suggested by Fraser \& Swinney (1986) as a tool to determine a
reasonable delay. The first minimum of the mutual information occured at
$\Delta$=13 (note, that henceforth we normalize $\Delta$ with the 10-days
long sampling rate) and a visual inspection of delay representations
supported the optimal
$\Delta$ to be between 10 and 15. The minimal embedding dimension was
estimated by the false nearest neighbour method proposed by Kennel et al.
(1992). False neighbours in the phase space occur when the reconstruction is
done with a $d_{\rm e}<d$ and points well-separated in higher dimensions get
closer due to projections into a lower dimension space. The idea is to find
that dimension in which the embedding results in marginal amount of false
neighbours, i.e. a ``good'' distribution of points is reached in the
reconstructed phase space. The routine {\tt false\_nearest} calculates the
fraction of false neighbours while changing $d_{\rm e}$ for a given distance
threshold. The results of this method should be handled carefully, because
in case of short datasets, it may underestimate the optimal dimension. For
short sets and higher dimensions one gets few neighbours simply because the
points sparsely occupy the embedding space. In our case $d_{\rm e}$=3 or 4
is suggested by this method, as the fraction of false neighbours decreases
drastically for larger embedding dimensions (even for 5 and 6). Therefore,
we adopted 10 to 15 as the reasonable range for $\Delta$ and 3, 4 for
$d_{\rm e}$.

\subsection{Recurrence plots}

\begin{figure}
\begin{center}
\leavevmode
\psfig{figure=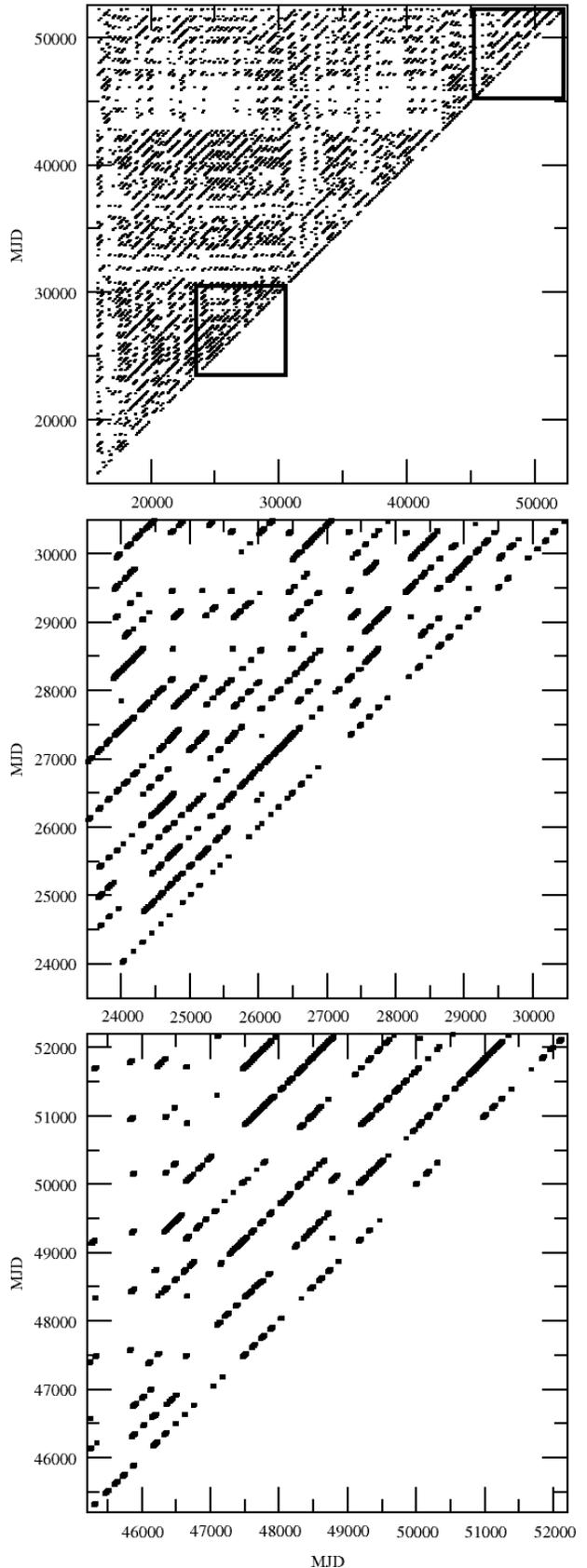,width=8.5cm}
\caption{The recurrence plot and two subregions with different
behaviour (marked with thick boxes in top panel).}
\end{center}
\label{fig9}
\end{figure}

To identify time resolved structures of the dataset, we have determined 
recurrence plots with various parameters (with the routine {\tt recurr}). 
It is a very simple tool: it scans the time series and marks each pair of
time indices $(i,j)$ whose corresponding delay vectors has distance 
$\leq \epsilon$ for a given $\epsilon$. Therefore, a black dot 
in the $(i,j)$-plane means closeness. If a periodic signal is examined, then
the points tend to occur both on the diagonal and parallel lines separated
from the main diagonal by integer multiplets of the period. These plots can
be very well used to detect and confine transients in a dataset.
Fortunately, the recurrence plot is not particularly sensitive to the choice
of embedding parameters and we found only marginal, almost invisible
differences when choosing embedding dimension and time delay within
reasonable ranges (3--4 for $d_{\rm e}$ and 10--15 for $\Delta$). We show
the recurrence plot of our data for $d_{\rm e}=3$ and $\Delta=13$ in (top
panel in Fig.\ 9), whose values were chosen without any special
consideration. In this Figure we have converted $(i,j)$ indices to MJD.

Several interesting conclusions can be drawn even from this very simple
plot. A few well determined transients occured between MJD 16300--17300,
30700--32500 and 42700--46300 (with a certain ambiguity of the boundaries).
The first one is probably caused by the lack of observations and thus
uncertain spline interpolation, but the other two are real as they are well
covered by data. Thus one has to be careful when analysing subsets that
include the transient regions. The periodicity is obvious from the dense
paralel lines in the plot. One can find large islands where the distance
between the parallel lines jumps to the doubled period. To emphasize such
details, we plot two subregions in the middle and bottom panels of Fig.\ 9.
The latter one
clearly corresponds to period-doubled region. By a close inspection of the
top panel, one can find similarly period-doubled regions, e.g., around
(20000,23000), (23000,41000) or (41000,41000) in $\sim1500\times1500$ wide
blocks. Furthermore, it is interesting that both two large transients were
similar as suggested by the clustered points around (30000,45000). It
follows that the transients must have occured around a particular small
region in the phase space, therefore, they were caused by the same process.

\subsection{Visualization of the trajectories}

\begin{figure*}
\begin{center}
\leavevmode
\psfig{figure=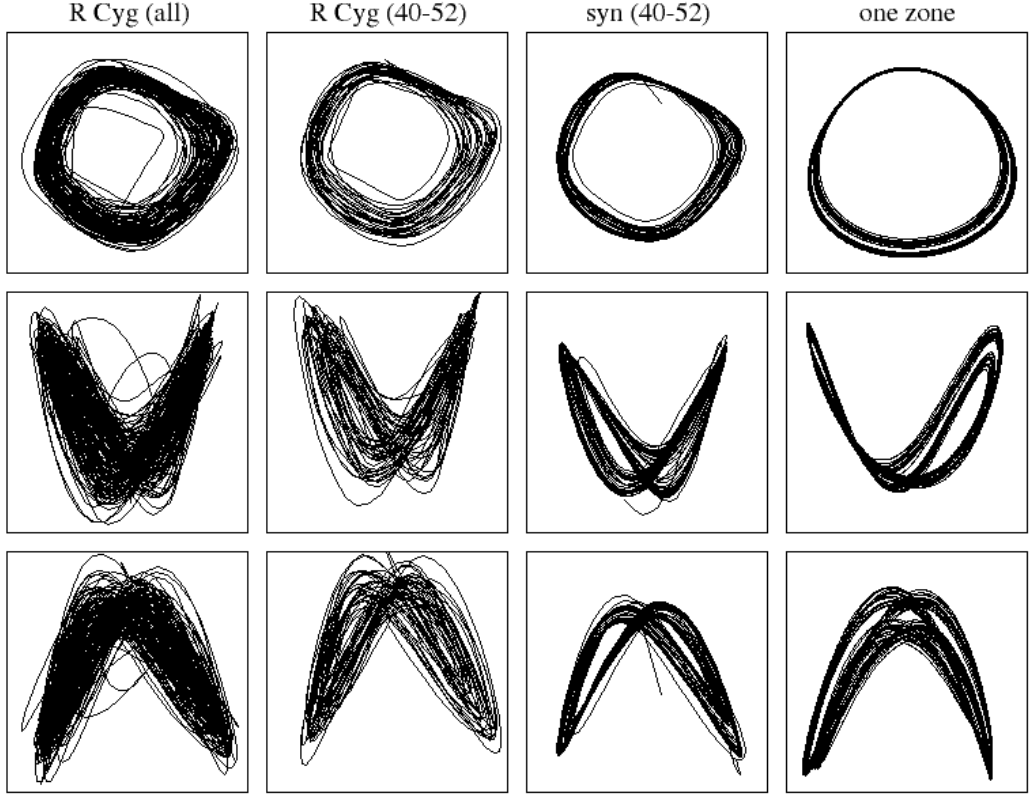,width=14cm}
\caption{BK projections ($d_{\rm e}=4$, $\Delta=10$). See text for further detais.}
\end{center}
\label{fig10}
\end{figure*}

To visualize the embedded state vectors $\mathbf{s}_{\rm n}$, 
we have generated the Broomhead-King (BK)
projections (Broomhead \& King 1986), which project onto the eigenvectors 
of the correlation matrix. In the TISEAN package, this can be 
done with the routine {\tt svd} (i.e. {\it singular value decomposition}).
We plot the most informative BK projections obtained from different 
data. The first column (R~Cyg (all)) is based on the whole $\{s(t_{\rm n})\}$,
with $d_{\rm e}=4$, $\Delta=10$. The second column (R~Cyg (40--52)),
with the same embedding parameters, contains the reconstructed
attractor from the last third
of the data, between MJD 40000 and 52200. Plots in the third column 
(syn (40--52)) are based on a synthetic dataset, which has been calculated by a locally
linear prediction using the routine {\tt nstep}. 
The determined local linear model can be iterated for an arbitrarily 
long time thus allowing the creation of synthetic datasets. 
In our case, the locally
linear approximation was iterated for 2400 points (i.e. 24000 days).
The reason for using only the last third of the data for the local linear
model is the fact, that this way we only approximate the unknown
map {\bf f} of the system (Buchler \& Koll\'ath 2001) with its local
Taylor expansion. For longer data the linear model oversmooths the
trajectories in the phase space thus blurring out the phase portrait of the
attractor.
The fourth column in Fig.\ 10 shows plots based on a simple one zone
model discussed in the next section. 

Our main conclusions are as follows. The embedding produces trajectories with 
remarkable structures. The global structure does not depend whether the 
whole set or some of its subsets are analysed. By iterating a locally linear
approximation of the map we could get a cleaner image of the attractor
which shows the period-doubling unambiguously. It is suggested
that the system switches back and forth between a period $2T$ orbit 
and chaotic state. The switch occurs in a located region of the 
phase state and there shows the light curve its transients.

\subsection{Lyapunov exponents and dimension, the correlation integral}

The exponential growth of infinitesimal perturbations from which the chaos
arises can be quantitatively characterized by the spectrum of Lyapunov
exponents (e.g. Kantz \& Schreiber, 1997 and references therein). 
If chaos is present in a dataset then one of the Lyapunov exponents
should be positive. We estimated this maximal exponent by computing

\begin{center}
$S(\epsilon,d_{\rm e},t)=\Bigg \langle {\rm ln} \Bigg ({1 \over | \mathscr{U}_{\rm n}|}
\sum\limits_{\mathbf{s}_{\rm n^\prime} \in \mathscr{U}_{\rm n}}
|s_{\rm n+t} - s_{\rm n^\prime+t}| \Bigg ) \Bigg \rangle_{\rm n} $  
\end{center} 

\noindent with the routine {\tt lyap\_k} ($\mathscr{U}_{\rm n}$ is the
$\epsilon$-neighbourhood of $\mathbf{s}_{\rm n}$, excluding $\mathbf{s}_{\rm
n}$ itself). If $S(\epsilon,d_{\rm e},t)$ shows a linear increase with the
same slope for all $d_{\rm e}$ larger than some $d_0$ and for a range of
$\epsilon$, then this slope is the estimate of the maximal exponent
$\lambda_1$. We calculeted $S(\epsilon,d_{\rm e},t)$ using the whole
$\{s(t_{\rm n})\}$ and plot the result in Fig.\ 11. Different branches 
corresponding to $d_{\rm e}$=3,4,5 and 6 and three values of $\epsilon$ 
show oscillations but are generally parallel. A global linear fit
gives a slope 0.0244 corresponding to $\lambda_1$=0.00244 d$^{-1}$. 
This roughy equals to that of determined for R~Sct ($\approx$0.0020 d$^{-1}$,
Buchler et al. 1996) and smaller by a factor of 2--3 as for AC~Her (Koll\'ath et al.
1998).

\begin{figure}
\begin{center}
\leavevmode
\psfig{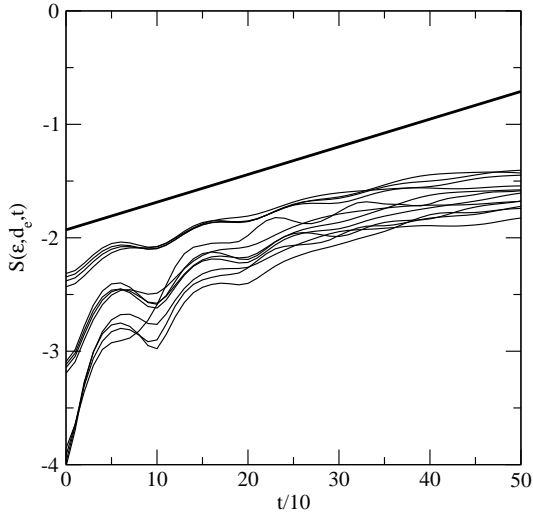}
\caption{Estimating the maximal Lyapunov exponent. The solid line is
the linear fit shifted vertically for clarity. The slope is 0.0244
implying $\lambda_1$=0.00244 d$^{-1}$.}
\end{center}
\label{fig11}
\end{figure}

The full Lyapunov spectrum was computed with the routine {\tt lyap\_spec}.
Its results should be considered as only preliminary, because the 
method, similarly to {\tt nstep}, employs local linear fits. It fits 
local Jacobians of the linearized dynamics and by multiplying them one by one
to different vectors in tangent space, emerges the trajectories.
The logarithms of rescaling factors are accumulated and their average give
the Lyapunov exponents (Sano \& Sawada 1985). 

In order to give an indication of the robustness of the Lyapunov spectrum
and Lyapunov dimension, we present their values for various embedding 
parameters in Table\ 2. In every cases, we find one positive exponent.
Thus, the attractor is clearly chaotic. In most cases, the absolute 
value of the second exponent is smaller than the first by a factor
of 3 to 5. An ideal flow should have $\lambda_2=0$ (Serre et al. 1996a), 
therefore our local fit is indeed only a rough approximation. 
Consequently, the determined Lyapunov dimensions scatter between 
2 and 3 with no definite tendency. Unfortunately, as noted, e.g.
in Serre et al. (1996a), the computation of all of the Lyapunov exponents
requires very long datasets. This is the other reason why our results 
are fairly uncertain. A global nonlinear fit would yield results that are
superior to ours presented here. We have to conclude that our
methods do not allow an accurate determination and a further study
should be devoted to {\it i)} a global flow reconstruction (Serre et al.
1996a) and {\it ii)} computation of more accurate Lyapunov exponents
and dimension. The latter one is important because these numbers are the main
quantitative parameters of a chaotic system.

\begin{table}
\begin{center}
\caption{Lyapunov exponents (in 10$^{-4}$ d$^{-1}$) and dimension.}
\begin{tabular}{|lr|rrrrrl|}
\hline
$d_{\rm e}$ & $\Delta$ & $\lambda_1$ & $\lambda_2$ & $\lambda_3$ & $\lambda_4$ & $\lambda_5$ & $d_{\rm L}$ \\
\hline
a) & & & & & & & \\
3 & 15 & 17 & $-$6 & $-$57 & & & 2.19\\
4 & 8 & 27 & $-$22 & $-$38 & $-$94 & & 2.12\\
4 & 10 & 28 & $-$9 & $-$24 & $-$114 & & 2.81\\
4 & 15 & 19 & $-$4 & $-$19 & $-$81 & & 2.78\\
5 & 5 & 31 & $-$30 & $-$44 & $-$72 & $-$126 & 2.02\\
\hline
b) & & & & & & & \\
3 & 15 & 26 & $-$7 & $-$67 & & & 2.30\\
4 & 8 & 38 & $-$16 & $-$45 & $-$97 & & 2.47\\
4 & 10 & 37 & $-$11 & $-$38 & $-$89 & & 2.70\\
5 & 5 & 30 & $-$19 & $-$35 & $-$65 & $-$182 & 2.30\\
\hline
\end{tabular}
\end{center}
$a)$ R~Cyg (40--52)\\
$b)$ syn (40--52)
\end{table}

Finally, the correlation dimension was estimated by the correlation sum 
$C(d_{\rm e},\epsilon)$ (Grassberger \& Procaccia 1983).
This method was applied by Canizzo et al. (1990) for three long-period 
variable stars. In those 
cases, Canizzo et al. concluded that the light curves can be modelled 
with a regular and a stochastic component thus rejecting the hypothetic
chaos. Buchler et al. (1996) argued against the use of correlation integral,
claiming that typically available data are too short for this purpose.
Nevertheless, we have computed the correlation sum to check its properties.
For this, the routine {\tt c2naive} was used which output was processed
by the routine {\tt c2d}. The latter routine computes 
${\rm d} \log C(\epsilon)/{\rm d} \log \epsilon$ for which one expects a wide
constant local minimum with different embedding dimensions (see detailed tests
presented in Canizzo et al. 1990).

\begin{figure}
\begin{center}
\leavevmode
\psfig{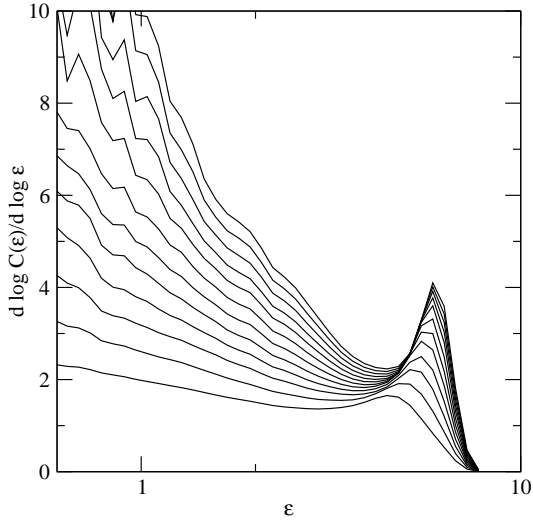}
\caption{Correlation dimension estimation from the correlation sum. 
Embedding dimensions 3 to 25 are shown for odd numbers. Slopes 
are determined by straight line fits to the log-log plot of 
the correlation sum. Note that horizontal axis is logarithmic.}
\end{center}
\label{fig12}
\end{figure}

We show the results of our computations in Fig.\ 12. This diagram is 
based on the whole $\{s(t_{\rm n})\}$. Although we could not find 
the expected linear scaling region, the presented behaviour slightly 
resembles that of the Lorenz-attractor as shown in Canizzo et al. (1990)
(see especially their Fig.\ 4b). The closely separated local minima
suggest a fractal dimension between 2.0 and 2.3 for the attractor,
though this estimate should be taken with caution. We conclude that
the analysis of the correlation sum yields results that are 
compatible with the previous ones.

\subsection{Comparison with a one zone model}

We compare the presented behaviour of R~Cygni with a simple
one zone model with stressed
period-doubling bifurcation. The similarity is provided by the 
reconstructed attractor. Although it has only marginal physical 
implications, the resemblance is quite illustrative. 

As a simple model for the dynamic of a pulsating white dwarf star, 
Goupil et al. (1988) used the following third order differential 
equation 

\begin{center}
$x^{\prime\prime\prime}+K x^{\prime\prime}+x^\prime + K \mu x (1+\beta x)=0$
\hskip1.5cm (1)
\end{center}

\noindent corresponding to a one zone model that is analogous to
Baker's one zone model (Baker 1966). Here $x$ is the radial displacement
from equilibrium. The control parameter is $\mu$, while $K=0.5$ and
$\beta=-0.5$ was fixed (see Goupil et al. 1988 for their meaning). When
$\mu$ is varied, the system first bifurcates from its stable fixed point to
a stable limit cycle (period $T$ orbit at $\mu=-1$), while at $\mu=-1.66$
the first period-doubling bifurcation occurs. The period $T$ orbit becomes
mildly unstable and a stable period $2T$ orbit exists. This was the main
point which reminded us to the case of R~Cygni and that is why we have
solved Eq.\ (1) numerically. During the fourth order Runge-Kutta numerical
integration we have added simple random perturbation that mimicked internal
perturbations of the system. To allow an easy comparison with R~Cygni, we
have rescaled the time to get 430 days for the period of the one zone model.

The solution is compared with the light curve and its iterated locally 
linear approximation in Fig.\ 13. The amplitude does not change
as much as in the original data, which is simply due to the fact, that
Eq.\ (1) gives the variations of the radius, not the luminosity. There
is, of course, some functional dependence between them, which is not
expected to change the very basic nature of the reconstructed attractor. 
The calculated maximum alternation, its disappearance and reappearance 
is similar to what characterizes R~Cygni, as has already been 
stressed by the BK projections. A further support is
given by the Fourier spectra of data shown in Fig.\ 14.
As discussed in Goupil et al. (1988), depending on how long and at
what time we observe the system, we see an oscillation whose trajectory
in phase space lies either on the period
$2T$ orbit or on the period $T$ orbit (then disappears the $f_0/2$ subharmonic),
or a mixed trajectory is observed switching over from one attractor
to the other. In this case, the temporal behaviour exhibits a change in
the alternation of small and large maxima. This largely reminds us of 
the light curve behaviour of R~Cyg.

\begin{figure}
\begin{center}
\leavevmode
\psfig{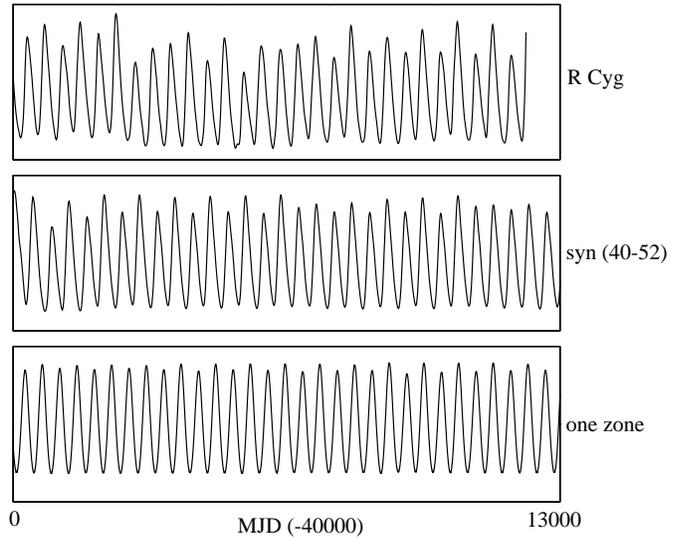}
\caption{A comparison of R~Cyg (40--52), syn (40--52) sets 
with the one zone model.}
\end{center}
\label{fig13}
\end{figure}

\begin{figure}
\begin{center}
\leavevmode
\psfig{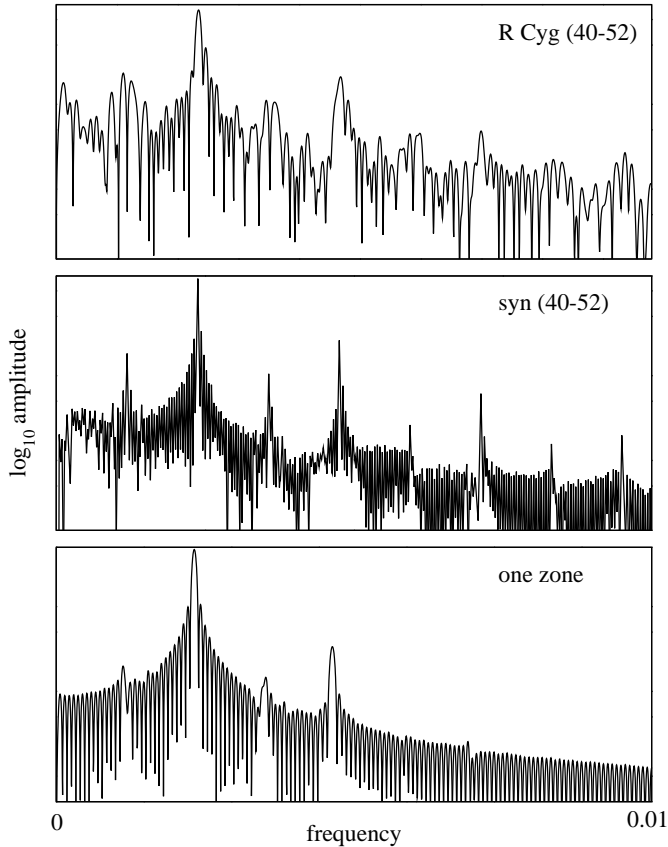}
\caption{Fourier spectra of the data plotted in Fig.\ 13.}
\end{center}
\label{fig14}
\end{figure}

\section{Conclusions}

Our study concentrated on the demonstration of the low-dimensional
chaotic behaviour in the light curve of a Mira star. For this, various
linear and nonlinear methods has been applied. While the $O-C$ diagram is
compatible with either stochastic (high dimensional random process) or 
chaotic (low-dimensional deterministic) dynamics, the frequency 
and time-frequency analyses raised the possibility of underlying chaos through
the presence of subharmonic components. The strongest pieces of evidence
came from the nonlinear analysis. Time delay embedding and different analyses
of embedded state vectors in the reconstructed phase space 
revealed the following results:

\noindent 1. The phase portraits show remarkably regular structures. The regularity 
does not depend whether the full dataset or some subsets are used for
embedding. The estimated minimum embedding dimension was 3 or 4 as suggested
the false nearest neighbours test. Other extensively used numerical 
methods favoured 4. 

\noindent 2. The long term behaviour is dominated with apparent 
period-doubling events. It can be interpreted as caused by a chaotic system 
which switches back and forth between a period $2T$ orbit and chaotic state.
The switch occurs in a compact region of the 
phase space in which the light curve exhibits 1000--1500-days long
transients.

\noindent 3. The Broomhead-King projections were used to visualize the 
phase portraits. Synthetic data were generated with a locally linear fit
to the attractor. They gave the clearest phase portraits 
showing unambiguously the period-doubling bifurcation. 

\noindent 4. Quantitative parameters, such as the Lyapunov exponents 
and dimension, were estimated, though with fairly large uncertainty. 
The estimated largest Lyapunov exponent is +0.0024 d$^{-1}$. This means
that the distances between the neighbouring trajectories increase
by $e$ during 417 days, i.e. we cannot predict the variation for 
longer time than one pulsation period. The Lyapunov spectrum was
computed with the locally linear approximation with various 
embedding parameters. In every cases we found one positive exponent.
In most case, the second one is close to 0 which is characteristic 
for ideal flows. The correlation sum, though rather inconclusively, 
suggests a fractal dimension 2.0--2.3 for the attractor.

\noindent 5. We have compared the analysed data with a simple one 
zone model generated by a third-order differential equation. It shows such
period-doubling bifurcation which is very similar to what is observed
for R~Cyg. The similarity of these two systems is our last argument
for low-dimensional chaos in the light curve of R~Cyg.

How can we compare our results with recent theoretical advancements in this
field of variable star research? Most importantly, presence of
low-dimensional chaos in a Mira star has been detected for the first time.
This means that despite the complex structure of an AGB star, the pulsations
in R~Cyg occur in a very dimension reduced phase space. Presently, we cannot
draw firm conclusions on the physical dimension of the system but our
results suggest 3 or 4. In the latter case, a similar consideration might be
applicable as for R~Sct by Buchler et al. (1996). In that case, they
concluded that the pulsations are the result of the nonlinear interaction of
two vibrationally normal modes of the star.  As soon as more sophisticated
nonlinear methods are used to analyse the light curve of R~Cyg, we will be
able to infer more physical implications on the nature of pulsations. A
further study will be devoted to a nonlinear global flow reconstrucion which
enables, e.g. computing accurate Lyapunov spectrum and dimension. Only then
it will be possible pinpointing the effective (physical) dimension of the
phase space.

The examination of publicly available databases (AFOEV, VSOLJ, BAAVSS)
revealed that there are at least
a few dozen or even more than one hundred Mira stars with continuous,
densely sampled visual light curves. R~Cygni is only one of them.
We list some Mira stars of similar period (400--470 days) 
with well-sampled light curves: T~Cas, R~And, R~Lep, R~Aur, U~Her, $\chi$~Cyg,
U~Cyg, V~Cyg, S~Cep, RZ~Peg, R~Cas and Y~Cas. A quick look at their
AFOEV and VSOLJ data reveals a few examples with interesting light curves
resembling that of R~Cyg. Another further task is to spot those 
stars which might be twins of R~Cyg and make an intercomparison between 
them. It is expected that the discovery of other 
similar stars may help to extend the observational base of a new 
discipline, the nonlinear asteroseismology (Buchler et al. 1996)
which extracts quantitative
information from the irregular light curves. It is, of course, 
still in its infancy and more theoretical as well as observational
efforts have to be undertaken. 

\begin{acknowledgements}
This research was supported by the ``Bolyai J\'anos'' Research Scholarship
of LLK from the Hungarian Academy of Sciences, FKFP Grant 0010/2001,
Hungarian OTKA Grants \#T032258 and \#T034615
and Szeged Observatory Foundation. 
We sincerely thank variable star observers of AFOEV, BAAVSS, 
and VSOLJ whose dedicated observations over a century made this
study possible. The computer service of the VSNET group is
also acknowledged. We are grateful to Dr. Z. Koll\'ath for providing
the TIFRAN software package and some useful comments that helped us to
improve our analysis. The NASA ADS Abstract Service was used to 
access data and references. This research has made use of Simbad Database
operated at CDS-Strasbourg, France.
\end{acknowledgements}

\end{document}